\documentclass[%
 reprint,
superscriptaddress,
 amsmath,amssymb,
 aps,
prb,
floatfix,
]{revtex4-2}

\usepackage{graphicx}% Include figure files
\usepackage{dcolumn}% Align table columns on decimal point
\usepackage{bm}% bold math
\usepackage[colorlinks,
           urlcolor=blue,
           linkcolor=blue,
           anchorcolor=blue,
           citecolor=blue
           ]{hyperref}% add hypertext capabilities
\usepackage{float}
\usepackage[dvipsnames]{xcolor}
\usepackage{soul}
\usepackage{siunitx}

\begin{document}

\preprint{APS/123-QED}

\title{Prediction of Hot Zone-center Optical Phonons in Laser Irradiated Molybdenum Disulfide with a Semiconductor Multitemperature Model}

% Force line breaks with \\

\author{Zherui Han}
 %\altaffiliation[Also at ]{Physics Department, XYZ University.}
 \affiliation{School of Mechanical Engineering and the Birck Nanotechnology Center,\\
Purdue University, West Lafayette, Indiana 47907-2088, USA}
\author{Peter Sokalski}
 \affiliation{Walker Department of Mechanical Engineering, The University of Texas at Austin, Austin, Texas 78712, USA}
%  \author{Rui Huang}
%  \affiliation{Department of Aerospace Engineering and Engineering Mechanics, University of Texas, Austin, Texas 78712, USA}
\author{Li Shi}
 \affiliation{Walker Department of Mechanical Engineering, The University of Texas at Austin, Austin, Texas 78712, USA}
\author{Xiulin Ruan}%
 \email{ruan@purdue.edu}
 \affiliation{School of Mechanical Engineering and the Birck Nanotechnology Center,\\
Purdue University, West Lafayette, Indiana 47907-2088, USA}

\date{\today}% It is always \today, today,

\begin{abstract}
Previous multitemperature model (MTM) resolving phonon temperatures at the polarization level and measurements have uncovered remarkable nonequilibrium among different phonon polarizations in laser irradiated graphene and metals. Here, we develop a semiconductor-specific MTM (SC-MTM) by including electron-hole pair generation, diffusion, and recombination, and show that a phonon polarization-level model does not yield observable polarization-based nonequilibrium in laser-irradiated molybdenum disulfide (MoS$_2$). In contrast, appreciable nonequilibrium is predicted between zone-center optical phonons and the other modes. The momentum-based nonequilibrium ratio is found to increase with decreasing laser spot size and interaction with a substrate. This finding is relevant to the understanding of the energy relaxation process in two-dimensional optoelectronic devices and Raman measurements of thermal transport. 

\end{abstract}

\maketitle

%\tableofcontents

Thermal relaxation of nonequilibrium charge and energy carriers is essential in operating semiconductor devices and laser processing of metals~\cite{Allenprl1987,STprl2017}. The energy cascade in such process is usually described as a hierarchical energy flow from an electrical or optical excitation to hot electrons, which are then coupled to optical phonons and finally decayed into the lattice. The efficiency of optoelectronic devices is increased by the so-called ``phonon bottleneck"~\cite{urayama2001phononbottleneck,nozik2002phononbottleneck,yang2017hotphonon} associated with the coupling between the hot electrons and phonons. This effect has been captured in previous two-temperature model (TTM) and its refined successors~\cite{Allenprl1987,qiu1992laser,MajumdarTTM,MomentumPRL2017,STprl2017} by assigning lumped temperatures to individual carrier subgroups. The TTM is also extended to consider semiconductors under photoexcitations~\cite{Shin2015ExtendedTTM}. Recent studies on thermalization of the lattice reveal that the nonequilibrium between the optical and acoustical phonon polarizations and that among different acoustical phonon polarizations can be even more pronounced than that between electrons and phonons~\cite{ramanprb2016,SeanNL2017}. The nonequilibrium inside the lattice phonon bath can give rise to nonthermal melting in devices~\cite{nonthermal2013} and inaccurate interpretation of thermal conductivity measurement based on Raman thermometry~\cite{ramanprb2016}. Past works have considered nonequilibrium between phonons with different frequency ranges or at different polarizations in the multitemperature model (MTM)~\cite{ramanprb2016,MTM} and the nonthermal lattice model (NLM)~\cite{NLM} that assign different temperatures to each phonon polarization or divide the phonons into low-frequency and high-frequency groups~\cite{coupleNL2017}. A polarization-based nonequilibrium has been considered in past studies of graphene~\cite{ramanprb2016,MTM,SeanNL2017} as well as a recent study of thin film molybdenum disulfide (MoS$_2$)~\cite{noneqMoS22021}.

The polarization-based nonequilibrium is a result of different coupling strengths between the hot electrons and different phonon polarizations, such as the three different acoustic phonon branches in a simple metal~\cite{NLM}. In addition, the out-of-plane polarized flexural (ZA) phonons in monolayer graphene show a peculiar quadratic dispersion and is subject to an additional restrictive electron-phonon scattering selection rule compared to that for the other two linear acoustic dispersions~\cite{MTM}. Meanwhile, the size confinement effects on in-plane phonons in nanosized graphene can modify the coupling strength between different phonon polarizations~\cite{coupleNL2017}.

In this work, we search for potential phonon nonequilibrium in MoS$_2$ thin films irradiated by a focused laser beam. The usual separation approach based on phonon energies or branches did not yield apparent nonequilibrium. Instead, we find that the zone-center optical phonons are hot by separation of phonons in both the energy and momentum spaces in our first-principles calculations of electron-phonon and phonon-phonon couplings. The calculated degree of nonequilibrium under the same experimental setup agrees well with a concurrent Raman experiment that reports a moderate nonequilibrium~\cite{peterExp}. The zone-center optical phonon temperature exceeds those of the other phonon modes increasingly at a reduced laser spot size and by enhanced cooling of the other modes via interface interaction between the MoS$_2$ thin layer and a substrate. While a prior study~\cite{MomentumPRL2017} has examined momentum-dependent electron-phonon coupling in multilayer tungsten diselenide (WSe$_2$) irradiated by ultrafast optical pulses, our results reveal that momentum-dependent electron-phonon and phonon-phonon coupling results in hot nonequilibrium phonons in both the energy and momentum spaces. Compared to recent theoretical predictions of zone-center and zone-boundary hot optical phonons in monolayer MoS$_2$~\cite{monolayerMoS22021jpcl} and MgB$_2$~\cite{MgB22020prl} during ultrafast pump-probe measurements, our theoretical calculation reveals zone-center hot optical phonons even in relatively thick MoS$_2$ flakes under steady-state focused laser heating and explains the finding from a concurrent Raman measurement~\cite{peterExp}. The current model has not considered excitons formed as bound electron-hole pairs, where the electrons are below the conduction band, and the holes are above the valence bands. The exciton binding energy of bulk MoS$_2$ might be close to room temperature. If we focus on a lattice temperature well above the room temperature with a high enough laser power, the thermal energy is higher than the exciton binding energy to break them into free electrons and holes in the conduction band and valence band, which are modeled here.
\begin{figure}[h]
    \centering
    \includegraphics[width=3.5in]{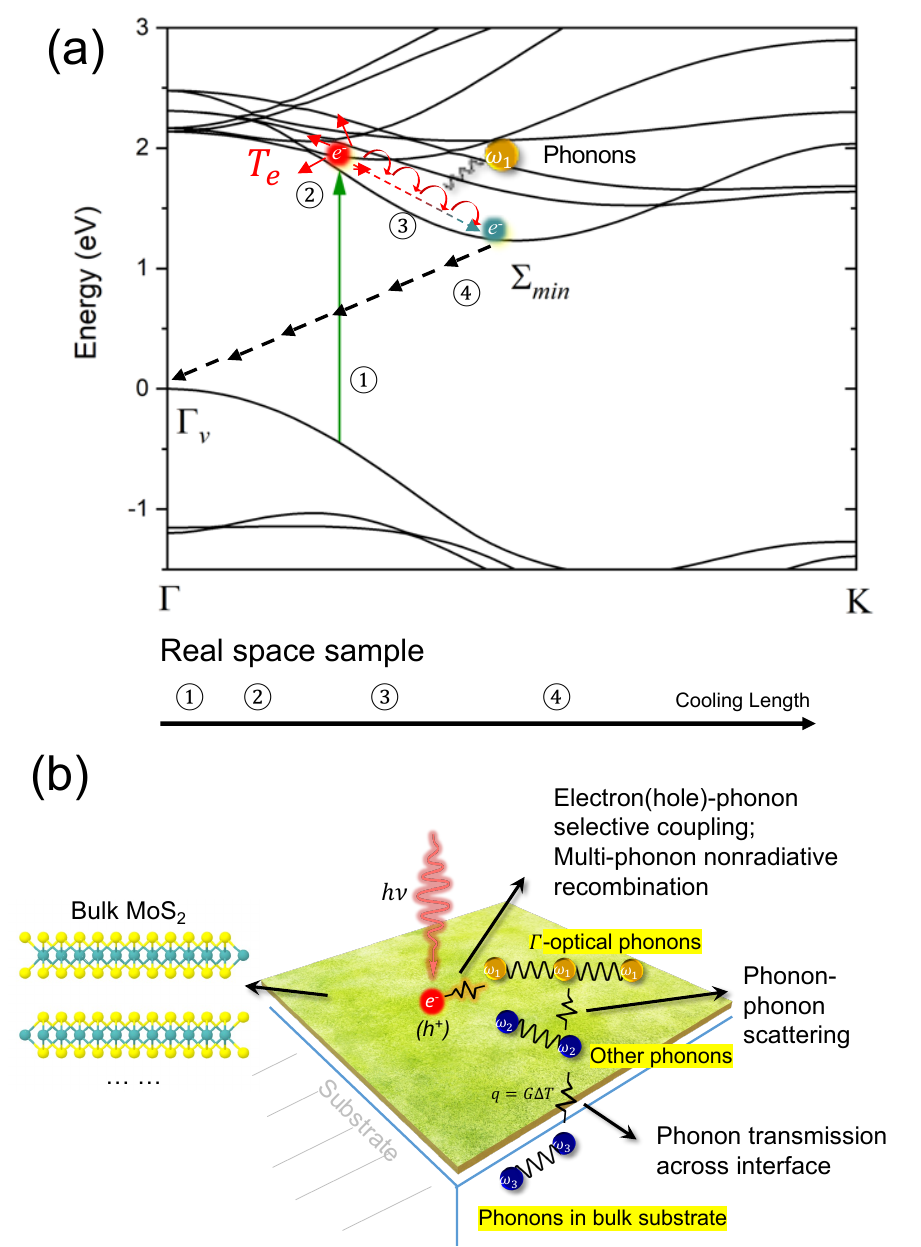}
    \caption{Schematic of energy pathway in localized laser heating on bulk MoS$_2$. (a) Hot electrons relaxation with different mechanisms mapped onto the calculated band structure of bulk MoS$_2$: \textcircled{1} photo-excitation, \textcircled{2} electron collisons, \textcircled{3} electron-phonon coupling, and \textcircled{4} multi-phonon recombination. In this plot, $\Gamma_v$ represents the valence band maximum and $\Sigma_{min}$ represents the conduction band minimum. The bottom real space illustration represents the associated different cooling lengths in the spatial domain. (b) Schematic showing the energy transfer in electron-phonon coupling \textcircled{3} and subsequent phonon-phonon scattering. $q$ is the heat flux through the lattice and $G$ is the thermal conductance between the material sample and the substrate. $\Gamma$-optical phonons (or zone-center optical phonons) are singled out in this work. Atomic structure in this figure is generated from a toolkit of Materials Project~\cite{MaterialsProject,Toolkit}.}
    \label{mechanism}
\end{figure}

Unlike metals, laser irradiation of semiconductors would create electron-hole pairs, which subsequently undergo hierarchical energy relaxation in time and/or spatial scales~\cite{Othonos1998Ultrafast,Shin2015ExtendedTTM}: quick thermalization of electrons (holes) by internal collisions, further energy relaxation by emission of phonons, and eventual energy decay by electron-hole recombination process. These processes are shown in Fig.~\ref{mechanism}(a) where we mark these relaxation processes onto calculated band structure. In the case of strong laser excitation, electrons and holes can be separated to have different temperatures and warrant separate treatment~\cite{Venkat2022ThreeTMSilicon}.
Figure~\ref{mechanism}(b) shows the energy dissipation through electron- (hole-)phonon interactions (EPI) and subsequent phonon-phonon scattering. Some phonon polarizations, usually the optical phonons, receive the majority of the dissipated energy and become overpopulated. The excess heat is then lost to the other phonon modes through phonon-phonon scatterings, including three-phonon and four-phonon processes. For a thin film material supported by a substrate, phonon-mediated heat transfer across the interface gives rise to an interfacial thermal conductance ($G_i$). Besides mode-dependent electron-phonon and phonon-phonon couplings in the thin layers, mode-dependent phonon coupling across the interface with the substrate phonons, usually characterized by large interface transmission coefficient for low-frequency acoustic vibrational modes~\cite{supportedgraphene2010}, is another mechanism that can give rise to nonequilibrium among different phonon modes in the thin layer.

\begin{figure*}[ht]
    \centering
    \includegraphics[width=7.1in]{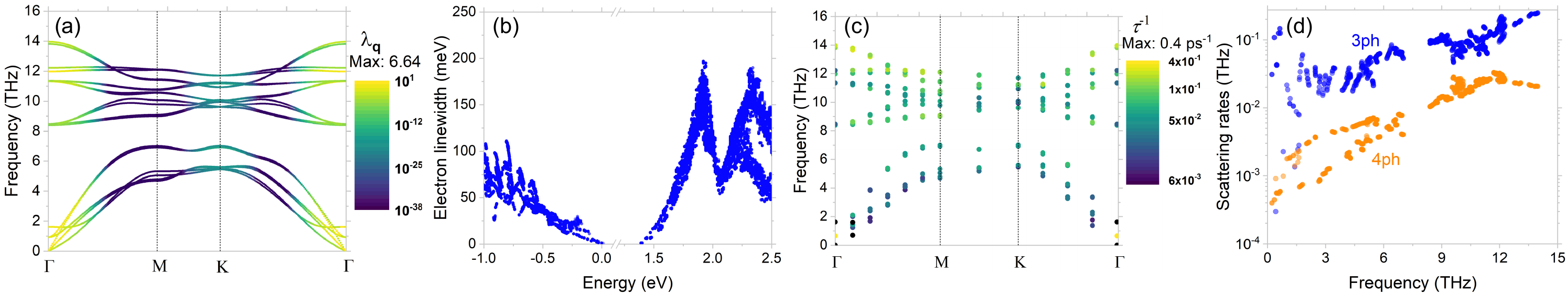}
    \caption{Electron-phonon coupling and phonon transport properties at 300~K in bulk MoS$_2$. (a) The electron-phonon coupling strength $\lambda_{\mathbf{q}}$ and its mapping onto the phonon dispersion curve. (b) Electron linewidth or imaginary part of its self-energy as a function of electron energy. This plot shows the contribution of electron-phonon scattering to the electronic linewidth. (c) Phonon-phonon scattering rates $\tau^{-1}$ and its mapping onto the phonon dispersion curve. (d) Anharmonic phonon scattering rates with contributions from three- ($\tau_{\mathrm{3ph}}^{-1}$, blue dots) and four-phonon scatterings ($\tau_{\mathrm{4ph}}^{-1}$, orange dots). In (a) and (c), the contour colorbar levels are in logarithm scale. }
    \label{e-ph}
\end{figure*}

Our first-principles calculations are performed in the framework of Density Functional Theory (DFT) as implemented in Vienna \textit{Ab initio} Simulation Package (VASP)~\cite{VASP1993} and we obtain the phononic structure using Density Functional Perturbation Theory (DFPT)~\cite{phonopy}. Interatomic force constants are calculated using finite difference method~\cite{shengbte}. Phonon-phonon coupling strength is then calculated using a package developed by some authors of the present work, \texttt{FourPhonon}~\cite{han2021fourphonon}. We use \texttt{QUANTUM-ESPRESSO}~\cite{QE} along with a modified \texttt{EPW} package~\cite{epw} to calculate electron-phonon coupling strength and further the electron scattering rates due to phonons. In this framework, coupling strength for a certain phonon mode ($\mathbf{q}$) is computed as $\lambda_{\mathbf{q}}=\frac{1}{N_{\rm F} \omega_{\mathbf{q}}} \displaystyle\sum_{\mathbf{k,k'}}|g_{\mathbf{k,k'}}^{\mathbf{q}}|^2\delta_{\mathbf{k}}\delta_{\mathbf{k'+q}}$~\cite{epw}, where $|g_{\mathbf{k,k'}}^{\mathbf{q}}|$ is the electron-phonon coupling matrix element involving electronic state $|\mathbf{k}\rangle,|\mathbf{k'+q}\rangle$ and phonon mode $|\mathbf{q}\rangle$ with frequency $\omega_{\mathbf{q}}$, $\delta$ is the Dirac delta function and $N_{\rm F}$ is the density of states. Further details on first-principles calculations are presented in Supplemental Materials~\cite{supply}.

An important question is how we should partition and regroup the energy carriers to represent the aforementioned origins of nonequilibrium phonons. For simplicity, we first consider suspended sample and set $G_i=0$ to examine the effects of mode-dependent electron-phonon and phonon-phonon coupling. We perform first-principles calculations to find spectral EPI and phonon scatterings. Figure~\ref{e-ph} shows our calculated mode-wise electron-phonon scattering and phonon-phonon scattering strengths that are mapped onto phonon dispersions in bulk MoS$_2$. We observe from Fig.~\ref{e-ph}(a) that electrons are mainly coupled with zone-center phonons and the coupling strength ($\lambda_{\mathbf{q}}$) for a certain mode specified by its momentum ($\mathbf{q}$) shows a strong dependence on the wavevector for each phonon branch. Additionally, some optical phonons with low frequencies and near the zone center still show strong coupling with electrons. This feature implies that grouping phonons only by their energy range or phonon branches may not be sufficient for this particular material. The calculated electron linewidth as a function of the electron energy is also shown in Fig.~\ref{e-ph}(b). In addition, we also obtain the electron-phonon coupling factor $G_{ep}$ for energy flow from electrons into different phonon groups. As shown by the calculated phonon-phonon scattering rates in Fig.~\ref{e-ph}(c,d), optical phonons have higher scattering rates that generally increases with increasing phonon frequency. A detailed analysis into different phonon scattering channels in Fig.~\ref{e-ph}(d) indicates that four-phonon scatterings ($\tau_{\mathrm{4ph}}^{-1}$) are generally one magnitude lower than three-phonon counterpart ($\tau_{\mathrm{3ph}}^{-1}$). Thus, in the present work we only calculate phonon coupling factors $G_{pp}$ based on three-phonon scattering rates. In our work, the coupling factors $G_{ep}$ and $G_{pp}$ are all obtained through first-principles calculations. 

In previous works~\cite{ramanprb2016,MTM}, the standard multitemperature model (MTM) resolves phonons at the polarization level without differentiating phonons from the same branch at different momentum spaces. Based on our mode-wise scattering rate calculation results for MoS$_2$, we specify five different equivalent temperatures for electrons, zone-center optical phonons, high-frequency non-zone-center optical phonons, low-frequency non-zone-center optical phonons, and acoustic phonons. In this way, the previous MTM~\cite{ramanprb2016,MTM} is extended to resolve phonons in both energy and momentum spaces. In addition, energy equation is modified to include recombination process and hole diffusion for semiconductor systems. We call this approach as semiconductor MTM (or SC-MTM):
\begin{equation}
\begin{aligned}
C_{e(h)} &\frac{\partial T_{e(h)}}{\partial t} =\nabla\left(\kappa_{e(h)} \nabla T_{e(h)}\right)+ r_{e(h)}\frac{Q}{h\nu}(h\nu-E_g) \\ 
 &-\sum_i G_{e(h) p, i}\left(T_{e(h)}- T_{p, i}\right)-\frac{3}{2}k_BT_{e(h)}n'_r,
\label{eeq}
\end{aligned}
\end{equation}

where the subscript $e(h)$ represents electrons (holes) and $(p,i)$ represents a certain phonon subgroup, $C$ denotes specific heat, $\kappa$ is the thermal conductivity, $T$ is temperature and $G$ is the coupling factor for each subgroup, $Q$ is external energy source, $E_g$ is the band gap, $h\nu$ is the photon energy, $\frac{3}{2}k_BT_{e(h)}$ is the thermal energy electrons (holes) have, $n'_r$ is the recombination rate. The ratio of effective mass is defined as $r_{e(h)}=\frac{m_{r,e(h)}^*}{m_{r,e}^*+m_{r,h}^*}$. In this formulation, the second term on the RHS represents the kinetic energy from photons in excess of the band gap gained by electron/hole, the third term is the cooling by electron- (hole-)phonon coupling, and the last term represents the thermal energy lost in recombination. Compared to metals, band gap should be subtracted when determining carrier temperatures.

For each phonon subgroup $i$, we have
\begin{equation}
\begin{aligned}
C_{p, i} \frac{\partial T_{p, i}}{\partial t}=& \nabla\left(\kappa_{p,i} \nabla T_{p,i}\right)+G_{e(h) p, i}\left(T_{e(h)}-T_{p, i}\right) \\
&+G_{p p, i}\left(T_{\text {lat}}-T_{p, i}\right) +\dot{q}_i.
\end{aligned}
\label{eph}
\end{equation}
In particular, we define a lattice reservoir $T_{\text {lat}}$ with which all phonon subgroups should exchange energy to satisfy energy conservation: $\sum G_{p p, i}\left(T_{\text {lat}}-T_{p, i}\right)=0$. By doing this, the phonon coupling factors are easily related to the phonon lifetime $\tau_{i}$ under relaxation time approximations (RTA): $G_{p p, i}=\frac{C_{p, i}}{\tau_{i}}$. Note that $\dot{q}_i$ is the fraction of the total recombination heating $\dot{q}$ that goes to group $i$ and we assume that all the recombination process is nonradiative. Considering the band gap energy is large compared to phonon energy, the nonradiative recombination will be a multi-phonon process. The most probable process would be depositing the energy to the zone-center optical phonons (detailed discussions on the above reasonings can be found in Supplymental Materials~\cite{supply}):
\begin{equation}
    \dot{q}(x)=\left(E_g+\frac{3}{2}k_BT_e\right)n'_r.
\end{equation}

Energy source $Q$ represents the absorbed Gaussian laser power density in an axisymmetric system and it reads $Q(r,z)=\alpha I(r,z)$ with $I(r,z)$ being the laser intensity profile~\cite{laserprl2016}:
\begin{equation}
I(r, z)=\frac{2 P\left(1-R\right)}{\pi w(z)^{2}} e^{-\frac{2 r^{2}}{w(z)^{2}}-\alpha z}.
\end{equation}
In the above equation, $P$ denotes the laser power, $R$ is the reflectance, $\alpha$ is the absorption coefficients of the sample, and $w(z)$ is the laser beam divergence relation at a certain depth $z$ (these parameters are discussed in Supplemental Materials~\cite{supply}). 

The recombination rate can be expressed as relaxation time form $n'_r \approx \frac{n}{\tau_n}$ where $\tau_n$ is the carrier recombination lifetime, $n$ is the excited carrier density. Thus, to solve the extended MTM energy equation Eq.~\ref{eeq} and~\ref{eph}, we need to determine the spatial carrier density $n$ by carrier diffusion equation~\cite{Streetman1975SolidStateElectronicDevices}:
\begin{equation}
    \frac{\partial n_{e(h)}}{\partial t}=D_{e(h)}\nabla^2 n_{e(h)}+n'_g-n'_r,
    \label{carrierdiffusion}
\end{equation}
where $D_{e(h)}$ is the diffusion coefficient of carriers, and $n'_g=\frac{Q}{h\nu}$ is the carrier generation rate. We estimate that electrons and holes in our material system have similar diffusion coefficients and effective masses~\cite{Bao2013MobilityMoS2,Cheiwchanchamnangij2012BandMoS2} and similar behaviors. Thus, in this specific work the solution of the above equations can be simplified to have just one equation where electrons and holes are lumped~\cite{supply}. Details and some discussions on electron/hole diffusion, carrier density profile and Fermi level are also presented in Supplemental Materials~\cite{supply} which include Ref.~\cite{Bao2013MobilityMoS2,Cheiwchanchamnangij2012BandMoS2,Kumar2013CarrierdynamicsMoS2,Yu2016PhotocurrentMoS2,Yuan2017CarrierdiffusionMoS2,Dagan2019CarrierMoS2}. Equations~\ref{eeq},~\ref{eph} and~\ref{carrierdiffusion} are solved numerically in real-space with axisymmetric boundary conditions using finite element methods in COMSOL Multiphysics~\cite{comsol}. 

\begin{figure}[h]
    \centering
    \includegraphics[width=3.3in]{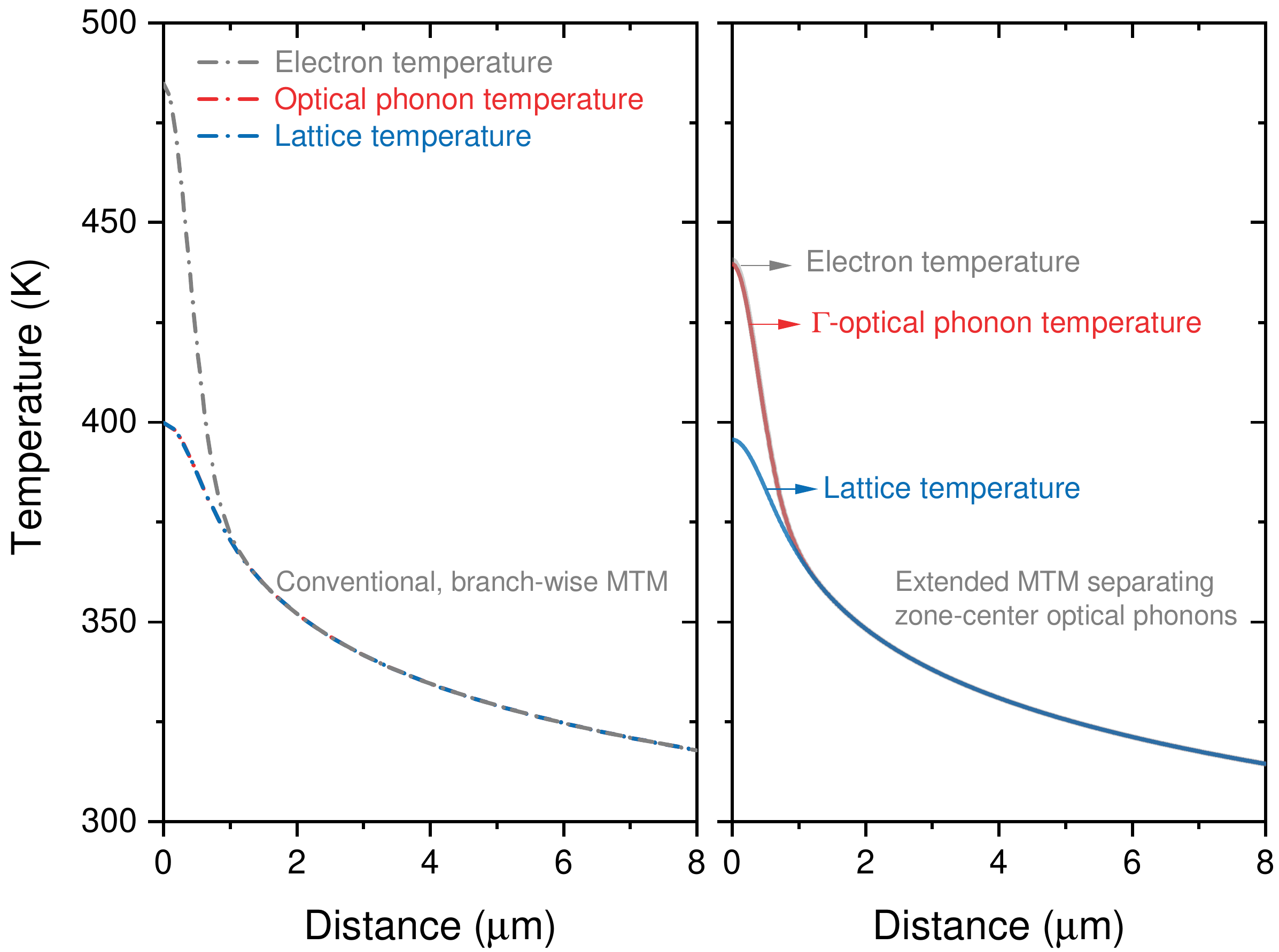}
    \caption{Extended semiconductor multitemperature model with both energy and momentum resolutions. The $x$ axis of this plot is the distance from the laser spot center. The suspended region of our sample is \SI{15}{\um} in radius and here we show the localized heating up to \SI{8}{\um}. The data is calculated under laser irradiation with the Gaussian radius $r=\SI{0.71}{\um}$ and power $P = 2.35~\rm mW$.}
    \label{temp}
\end{figure}

Figure~\ref{temp} compares the calculation results from the branch-wise SC-MTM and the present momentum-resolved SC-MTM model. The branch-wise MTM assigns two different average phonon temperatures, one for the optical branches and the other for the acoustical branches. While the previous branch-wise MTM can reveal phonon nonequilibrium in metals and graphene~\cite{ramanprb2016,MTM,NLM,coupleNL2017}, its semiconductor modification cannot produce any observable phonon nonequilibrium in bulk MoS$_2$, as shown in Fig.~\ref{temp}(a). In comparison, the momentum-resolved SC-MTM model reveals apparently higher temperatures for the zone-center optical phonons than $T_{\rm lat}$. This result suggests that the zone-center optical phonons in bulk MoS$_2$ are coupled preferentially to electrons and are not effectively cooled down by the lattice. Based on this result, phonon nonequilibrium in laser irradiated MoS$_2$ mainly occurs in the momentum space between the zone center modes and other modes, instead of between phonons of different polarizations. Essentially, the energy flow bottleneck is mainly in the momentum space instead of across the different energy ranges or polarizations. Electron temperature (shown in gray lines) is also higher in the eyes of branch-wise SC-MTM than predicted by momentum-resolved SC-MTM. Electrons are at equilibrium with zone-center optical phonons as revealed by our extended model.

\begin{figure}[h]
    \centering
    \includegraphics[width=3.5in]{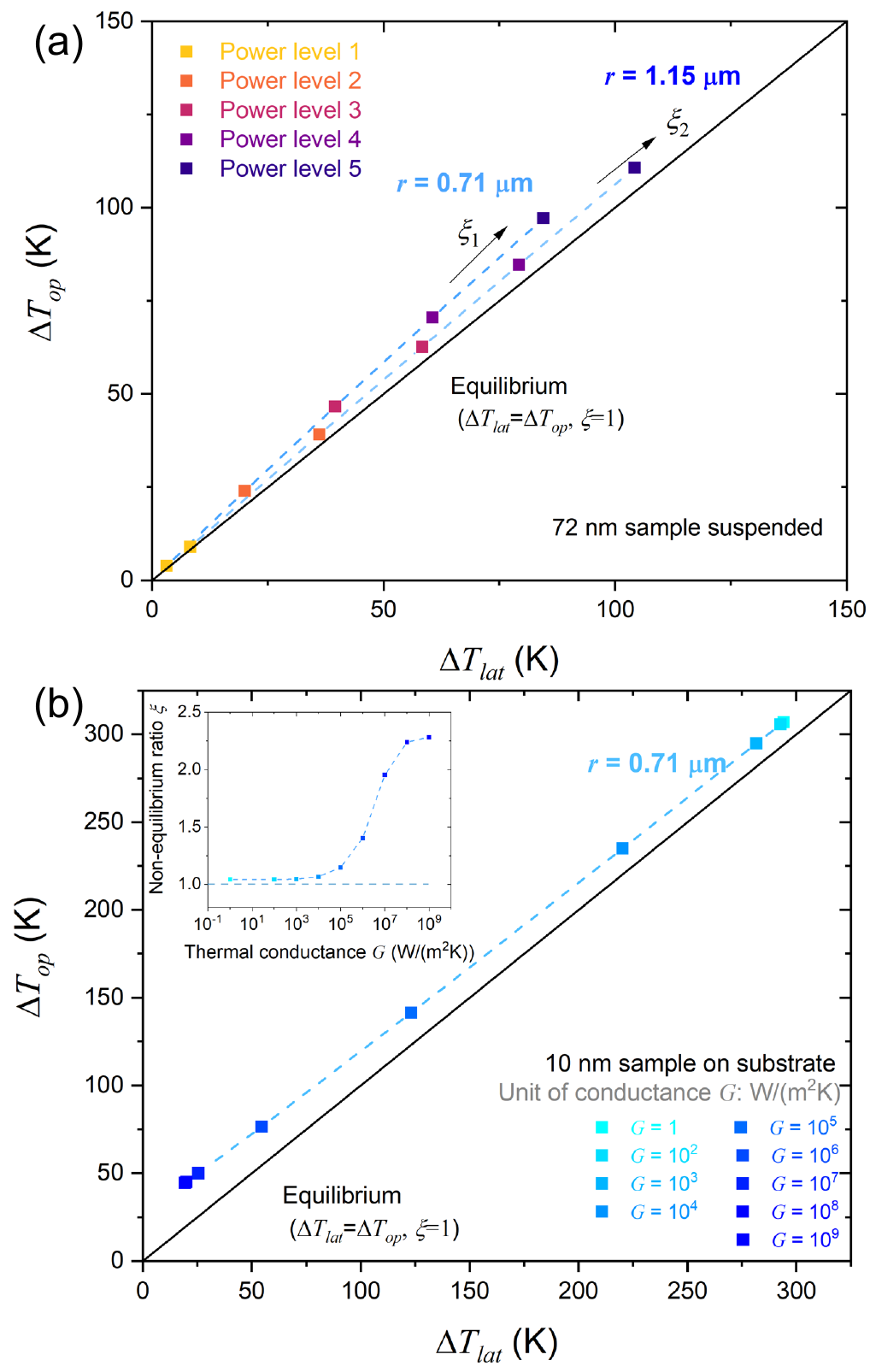}
    \caption{Effect of laser spot size and substrate coupling on the nonequilibrium ratio $\xi$. $\xi$ is interpreted as the slope of line passing through the origin. (a) The hottest phonon temperature rise with respect to the lattice temperature rise at increasing laser power levels 1 to 5 for suspended MoS$_2$ that has a thickness of 72~nm. The slope yields $\xi$, and $r$ is the laser spot size. The solid black line with unity slope indicates the equilibrium state. (b) The hottest phonon temperature rise with respect to the lattice temperature rise for supported MoS$_2$ that has a thickness of 10~nm at different thermal interface conductance, a fixed laser radius $r=\SI{0.71}{\um}$ and power $P = 2.35~\rm mW$. The inset shows the calculated nonequilibrium ratio $\xi$ as a function of conductance $G_i$. }
    \label{ratio}
\end{figure}

We define a nonequilibrium ratio as $\xi=\Delta T_{\rm h}/\Delta T_{\rm lat}$ between the temperature rise of the hottest group ($\Delta T_{\rm h}$) and that of the lattice ($\Delta T_{\rm lat}$). In the case of bulk MoS$_2$, $\Delta T_{\rm h}$ is the temperature rise ($\Delta T_{\rm op}$) of the zone-center optical phonons. In Fig.~\ref{ratio}(a) that has accounted for the effect of the laser-induced temperature rise, this ratio is obtained as the slope of the calculated $\Delta T_{\rm op}$ versus $\Delta T_{\rm lat}$ at different laser power and each laser spot size for suspended sample. Though the change of nonequilibrium with respect to laser power is marginal in this system (indicated by the almost linear relation in Fig.~\ref{ratio}(a)), the numerical solutions do show a decrease of nonequilibrium with higher laser power level. This power dependence is in agreement with and explained in the prior study of graphene~\cite{SeanNL2017}. Our simulated nonequilibrium ratio under the same experimental condition is 1.15 when $r=\SI{0.71}{\um}$, in good agreement with Ref.~\cite{peterExp}. Compared to the degree of phonon nonequilibrium observed in graphene ($\xi>2$)~\cite{ramanprb2016,SeanNL2017}, this reduced ratio is understandable as bulk MoS$_2$ has lower thermal conductivity and generally stronger phonon scatterings than graphene. Additionally, the ZA mode in graphene couples weakly with other phonon modes and essentially cools down the lattice. When the Gaussian laser beam radius decreases from \SI{1.15}{\um} to \SI{0.71}{\um} and approaches the thermalization length of hot zone-center phonons, the zone-center phonons are not able to thermalize with the lattice within the laser spot, resulting in a higher $\xi$ at a smaller laser spot size.

We further examine the effect of substrate phonon coupling on the nonequilibrium ratio, as the findings will have practical implications for electronic devices made with supported MoS$_2$ layers~\cite{TR2022AMI}. Thus, we choose a model sample with a thickness of 10~nm on SiO$_2$ substrate in this case. Acoustic phonon modes of SiO$_2$ lie in the same frequency range as MoS$_2$ ($\sim$5~THz). Since high-group-velocity acoustic phonons usually have the dominant contribution to the interface conductance and considering the fact that the low-frequency optical phonons in bulk MoS$_2$ are in near equilibrium with the lattice, in our model the interface conductance term is only added to acoustic phonon group, where the energy outflow to the substrate is proportional to the temperature difference between the acoustic group $(p,a)$ and the substrate ($T_{\rm sub}$):
\begin{equation}
\begin{aligned}
C_{p, a} \frac{\partial T_{p, a}}{\partial t}=\nabla\left(\kappa_{p, a} \nabla T_{p, a}\right)+G_{e p, a}\left(T_{e}-T_{p, a}\right)+ \\
G_{p p, a}\left(T_{\mathrm{lat}}-T_{p, a}\right)-G_{i}\left(T_{p, a}-T_{\text {sub }}\right).
\end{aligned}
\label{eq.substrate}
\end{equation}
As shown in Fig.~\ref{ratio}(b), an increase in the interface conductance $G_i$ would decrease the overall lattice temperature rise $\Delta T_{\rm lat}$ more than $\Delta T_{\rm op}$ for a given laser power level and laser spot size, because $\Delta T_{\rm lat}$ is cooled directly by interface heat transport mainly caused by interface coupling of acoustic phonons. Consequently, the nonequilibrium ratio $\xi$ increases with $G_i$. 

These results show that the conventional phonon branch-wise multitemperature model (MTM) revised for semiconductors (SC-MTM) does not yield any observable phonon nonequilibrium in laser-irradiated MoS$_2$. However, when we extend the model to have both energy and momentum dimensions, a moderate nonequilibrium is produced and can explain the experimental data in Ref.~\cite{peterExp}. This implies that thermal equilibrium is well established between phonon groups having different energy states, but not between phonons with different momenta even in the same branch. Only the extended SC-MTM that treats the zone-center phonon separately, while not the conventional phonon branch-wise SC-MTM, can predict the phonon nonequilibrium that is observed in concurrent experiments. Based on this finding, whether phonons should be resolved in both energy and momentum space should be based on first-principles calculations of electron-phonon and phonon-phonon interactions. The degree of hot zone-center optical phonons is increased with reduced laser spot size and substrate coupling of acoustic modes. These findings provide detailed insights on phonon nonequilibrium phenomena in optically or electrically excited functional materials and devices.

\begin{acknowledgments}
X. R., Z. H., and L. S., P. S. were supported by two collaborating grants (No.~2015946 and No.~2015954) of the U.S. National Science Foundation. Simulations were performed at the Rosen Center for Advanced Computing (RCAC) of Purdue University. 

\end{acknowledgments}

% \nocite{*}

\bibliography{Reference}% Produces the bibliography via BibTeX.

\end{document}

\title{Supplemental Material for ``Prediction of Hot Zone-center Optical Phonons in Laser Irradiated Molybdenum Disulfide with a Semiconductor Multitemperature Model"}% Force line breaks with \\

\author{Zherui Han}
 %\altaffiliation[Also at ]{Physics Department, XYZ University.}
 \affiliation{School of Mechanical Engineering and the Birck Nanotechnology Center,\\
Purdue University, West Lafayette, Indiana 47907-2088, USA}
\author{Peter Sokalski}
 \affiliation{Walker Department of Mechanical Engineering, The University of Texas at Austin, Austin, Texas 78712, USA}
%  \author{Rui Huang}
%  \affiliation{Department of Aerospace Engineering and Engineering Mechanics, University of Texas, Austin, Texas 78712, USA}
\author{Li Shi}
 \affiliation{Walker Department of Mechanical Engineering, The University of Texas at Austin, Austin, Texas 78712, USA}
\author{Xiulin Ruan}%
 \email{ruan@purdue.edu}
 \affiliation{School of Mechanical Engineering and the Birck Nanotechnology Center,\\
Purdue University, West Lafayette, Indiana 47907-2088, USA}

\date{\today}% It is always \today, today,

\maketitle
\tableofcontents

\section{Carrier density profile, hole diffusion and Fermi level}

If $D_e$ differs significantly from $D_h$, electrons and holes can evolve differently~\cite{Venkat2022ThreeTMSilicon}. In the case of bulk MoS$_2$, field effect mobility measurements give quite similar results~\cite{Bao2013MobilityMoS2}: for a sample with 50~nm thickness, electron mobility is 470~cm$^2$/Vs while hole mobility is 480~cm$^2$/Vs. This is consistent with a report where their effective masses are calculated to be similar~\cite{Cheiwchanchamnangij2012BandMoS2}. Considering the Einstein relation ($D=\frac{\mu k_B T}{q}$) and low carrier density shown later, we estimate that electrons and holes in our material system have similar diffusion coefficients and similar behaviors. Thus, in this specific work the solution of the above equations can be simplified to have just one equation where electrons and holes are lumped.

When electrons are pumped, Fermi level in semiconductors should split into two quasi-Fermi levels ($E_f^n,E_f^h$) and shift with increasing carrier density. In some way, for intrinsic semiconductors this is effectively ``photo-doping" with equal number of electrons and holes. The carrier concentration equations give us such relations~\cite{Streetman1975SolidStateElectronicDevices}:
\begin{align}
    E_f^n - E_i &= k_B T \ln{\frac{n_e(x)}{n_0}}, \\
    E_f^p - E_i &= -k_B T \ln{\frac{n_h(x)}{n_0}},
\end{align}
where $E_i$ is at the middle of the band gap.

\begin{figure}[!htb]
    \centering
    \includegraphics[width=6in]{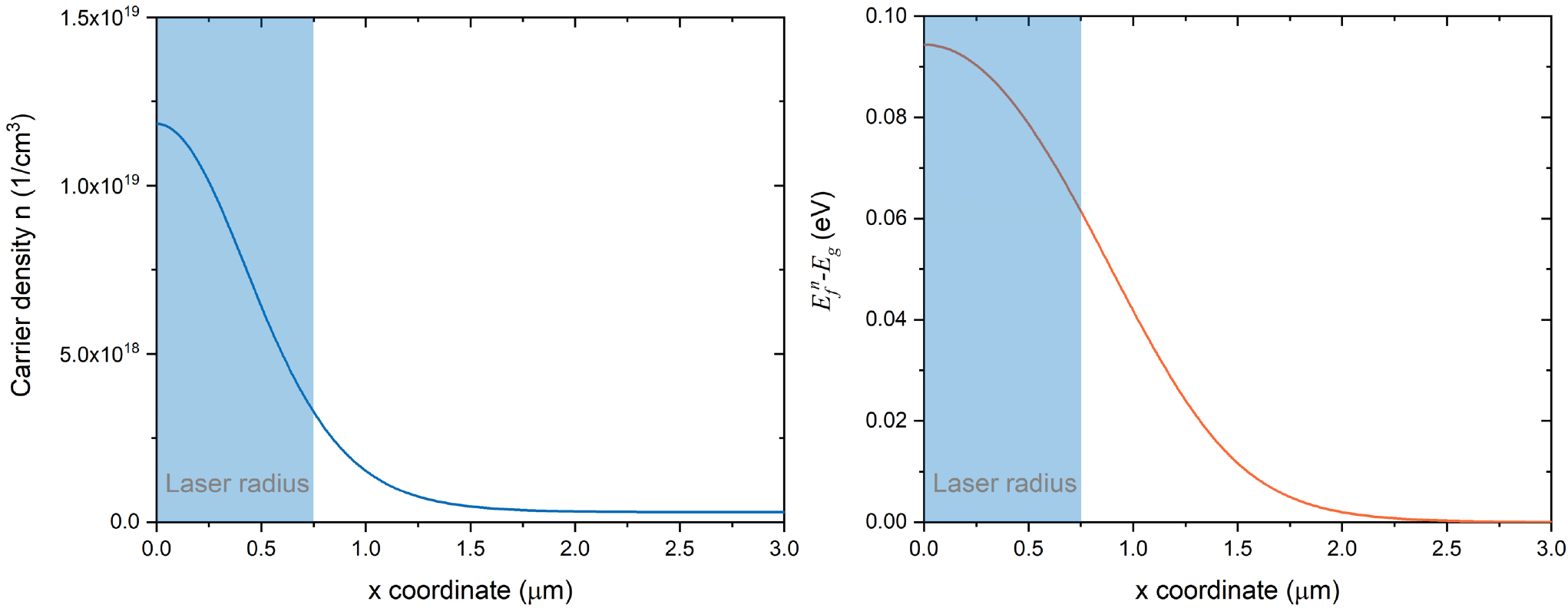}
    \caption{Surface spatial profile of carrier density (left) and Fermi level shift (right) at steady-state due to photoexcitation. Shaded area is laser radius of focused laser beam. Laser parameter is radius $w=0.71~\rm \mu m$ and power $P = 2.35~\rm mW$.}
    \label{carrierdensity}
\end{figure}

Recall that carrier diffusion equation is now reduced to 
\begin{equation}
    \frac{\partial n_{e}}{\partial t}=D_{e}\nabla^2 n_{e}+n'_g-n'_r.
    \label{carrierdiffusion}
\end{equation}
Using the experimentally determined parameters~\cite{Kumar2013CarrierdynamicsMoS2}, we have $D_e=4.2~\rm cm^2/s$ and $\tau_n=180~\rm ps$. The carrier density at equilibrium is $n_0\sim 3\times10^{17}\rm cm^{-3}$~\cite{Dagan2019CarrierMoS2} and $n'_r \approx (n_e-n_0)/\tau_n$. We can then solve for carrier density and its spatial profile. The simulated carrier density and Fermi level shift at the surface of MoS$_2$ are presented in Fig.~\ref{carrierdensity}. The largest Fermi level change (at the center of laser spot and at the surface of the sample) is less than 0.1~eV and the average across the thickness is less than 0.05~eV.  We then conclude that in our parallel experiment effort~\cite{peterExp} carrier injection is of low-level and Fermi level shift is marginal. As also argued in Ref.~\cite{STprl2017} of the manuscript, some change of Fermi level does not have large effect on the computed $e$-ph coupling rates. Our evaluation of electron-phonon coupling strength should capture the real energy relaxation dynamics in this semiconductor.

\section{Recombination heating and its deposition to phonon group}

In this section, we justify our treatments on recombination process in the main text. We first assume that all the recombination process is nonradiative and the energy goes to the lattice. This assumption is supported by two arguments: (1) the quantum efficiency of multilayer MoS$_2$ is less than 7\%~\cite{Yu2016PhotocurrentMoS2}; (2) bulk MoS$_2$ has indirect band gap and the radiative recombination is restricted by conservation of momentum~\cite{Yuan2017CarrierdiffusionMoS2}.

Our next task is to estimate how $\dot{q}$ is deposited into the lattice. Indirect band gap is $E_g\sim 1.3$~eV for bulk MoS$_2$ while the largest phonon energy is around 0.06~eV. The most likely channel is by emitting about 20 high-frequency phonons (emitting greater number of phonons is higher-order process). The indirect nature of bulk MoS$_2$ (CBM around the middle of $\Gamma$ to K path) indicates that these phonons have small momentum: considering the number of emitted phonons (around 20), their momenta are well within our simulation grid and can be lumped into zone-center phonons. Thus, we assume that the majority of the recombination heating $\dot{q}$ is deposited into zone-center optical phonons:
\begin{equation}
    C_{p, zc} \frac{\partial T_{p, zc}}{\partial t}= \nabla\left(\kappa_{p,zc} \nabla T_{p,zc}\right)+G_{e p, zc}\left(T_{e}-T_{p, zc}\right)+G_{p p, zc}\left(T_{\text {lat}}-T_{p, zc}\right)+\dot{q}.
    \label{phonon-recombine}
\end{equation}

\section{First-principles calculations for electron-phonon and phonon-phonon couplings}
In this section, we cover the computational details in our first-principles calculations to obtain input parameters in the extended MTM.

All calculations are done using Density Functional Theory (DFT) or Density Functional Perturbation Theory (DFPT). For structural optimization, we use $k$-grid of $12\times12\times3$ with 520~eV plane wave energy cutoff. To describe van der Waals (vdW) interactions in bulk MoS$_2$, we employ a DFT-D3 correction method~\cite{DFT-D3}. For electron-phonon interactions, all calculations are done using the QUANTUM-ESPRESSO package~\cite{QE} with Perdew-Burke-Ernzerhof exchange-correlation functional~\cite{prl1996GGA}, norm-conserving pseudopotentials~\cite{nc-potential}, and a kinetic energy cutoff of 50~Ry. The EPW package~\cite{EPW} is employed to perform Wannier function interpolation for the e-ph coupling matrix elements. The related quantities are calculated first on a coarse grid of $8 \times 8 \times 2$ and then Wannier interpolated into a fine gird of $32\times32\times8$ for electron linewidth, and $16\times16\times4$ $q$-mesh for phonon self energy. The electronic integration over the Brillouin Zone (BZ) is approximated by the Gaussian smearing of 0.02 eV for the self-consistent calculations.

For phonon-phonon interactions in bulk MoS$_2$, we employ the VASP package~\cite{VASP1993} and use Perdew-Burke-Ernzerhof (PBE) parameterization of the generalized gradient approximation (GGA) for exchange and correlation functionals~\cite{prl1996GGA}. The plane wave cutoff is 520~eV. We also construct $4 \times 4 \times 1$ supercells and use $3 \times 3 \times 2$ $k$-mesh to calculate interatomic force constants (IFCs) and consider 0.5~nm cutoff and the second nearest neighboring atoms for the third-order IFCs and fourth-order IFCs, respectively. The BZ is discretized by $12 \times 12 \times 4$ $q$-mesh to evaluate three-phonon scattering rates using ShengBTE package~\cite{shengbte} and four-phonon scattering rates using the FourPhonon tool~\cite{han2021fourphonon}. The mode-resolved thermal conductivity is computed using a slightly modified version of the FourPhonon package. The broadening factor for the three-phonon calculations is unity. In this work, the thermal expansion and phonon renormalization effects are not included.

\section{Temperature dependence of coupling parameters in the extended MTM}

In the extended MTM, the input parameters should in principle all be temperature-dependent but their detailed mechanisms of temperature dependence are different. For electron relaxation, the coupling factor is computed as $G_{ep,i}=\frac{C_e}{\tau_{e,i}}$ where $C_e$ is the electron specific heat and  $\tau_{e,i}^{-1}$ is the electron scattering rates (or electron linewidth) due to phonon subgroup $i$. Electron specific heat $C_e$ is computed as the derivative of electron energy with respect to the electron temperature $T_e$:
\begin{equation}
C_{\mathrm{e}}(T_e)=\int_{-\infty}^{\infty} \rho(\varepsilon) \varepsilon \frac{\partial f\left(\varepsilon-\mu,T_{\mathrm{e}}\right)}{\partial T_{\mathrm{e}}} d \varepsilon,
\end{equation}
where $\rho$ is the electron density of states, $f$ is the Fermi-Dirac distribution, $\mu$ is the electronic chemical potential and $\varepsilon$ is the electron energy.
In principle, $G_{ep,i}$ and $\tau_{e,i}^{-1}$ have complicated temperature dependence on both electron temperature and phonon temperature, as higher electron temperature means more electrons are excited into higher energy states, and higher phonon temperature means greater scattering with electrons. In Ref.~\cite{ramanprb2016}, this is captured by several third-order polynomials with two independent variables $T_e$, $T_{p,i}$ and a fitting into experimental data at some temperatures. In Ref.~\cite{MgB22020prl}, the authors proposed a more convenient way such that electron specific heat $C_e$ has a linear dependence on $T_e$ and it makes $G_{ep,i}$ insensitive on $T_e$ altogether. This consideration simplifies the temperature dependence here and in this work we can express $G_{ep,i}(T)=G_{ep,i}(T_{p,i})$. We can then evaluate electron scattering rates for an electron state $\mathbf{k}$ due to phonon group $i$ at a phonon group temperature:

\begin{equation}
\begin{aligned}
\frac{1}{\tau_{\mathbf{k},i}}(T)= \sum_{\mathbf{q}\in i}\left|g_{\mathbf{k,q}}\right|^{2}\left\{\left[\left(f_{ \mathbf{k+q}}+n_{\mathbf{q}}\right) \delta\left(\varepsilon_{ \mathbf{k+q}}-\varepsilon_{\mathbf{k}}-\hbar\omega_{\mathbf{q}}\right)\right]\right.
\left.+\left[\left(1+n_{\mathbf{q}}-f_{\mathbf{k+q}}\right) \delta\left(\varepsilon_{\mathbf{k+q}}-\varepsilon_{\mathbf{k}}+\hbar\omega_{\mathbf{q}}\right)\right]\right\},
\end{aligned}
\label{ep-tau}
\end{equation}
where we perform summation over phonon modes $\mathbf{q}$ in subgroup $i$, and $f$ and $n$ are Fermi-Dirac and Bose-Einstein distributions, respectively. $\varepsilon$ is the corresponding electronic state energy and $\omega$ is the phonon frequency.

\begin{figure}[h]
    \centering
    \includegraphics[width=6.5in]{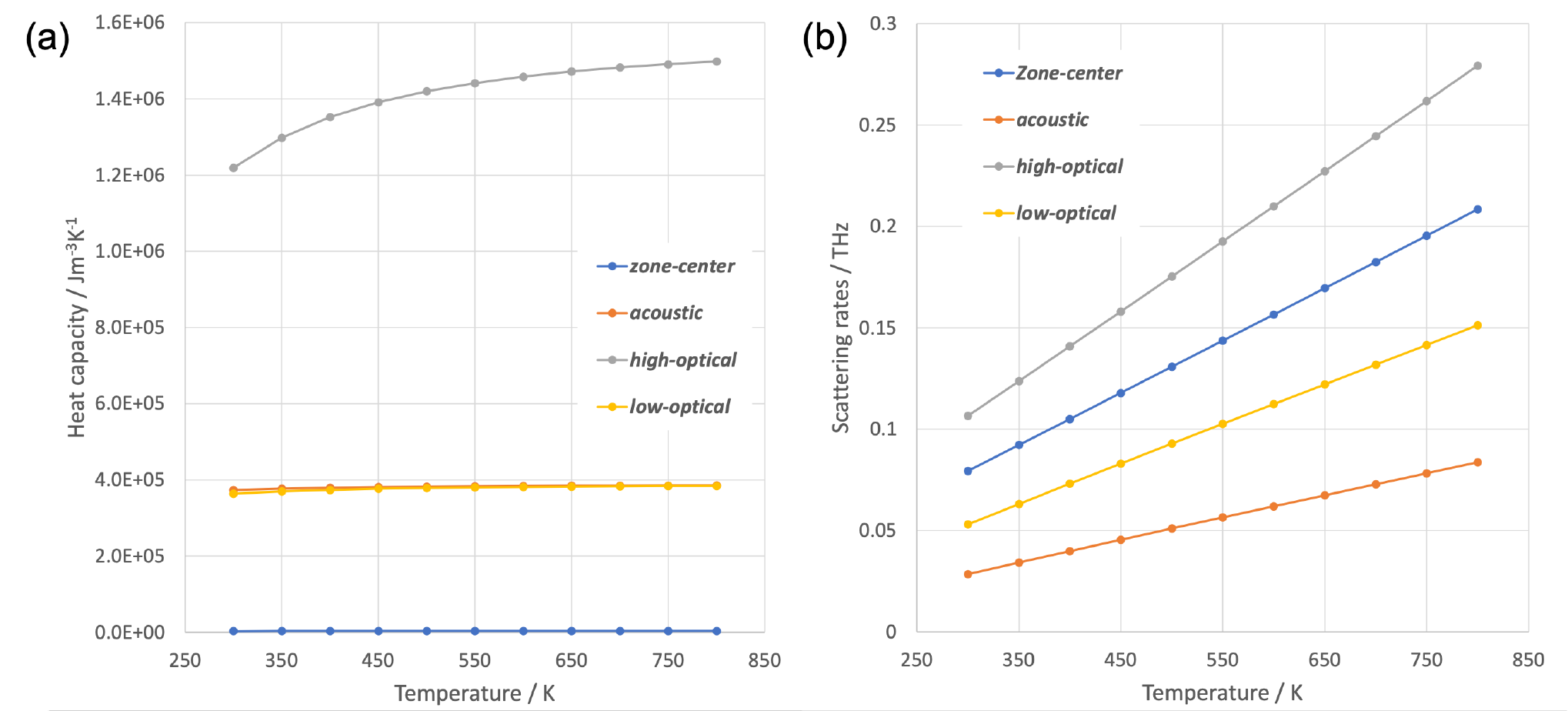}
    \caption{Temperature-dependent phonon properties of four subgroups in this study: (a) specific heat and (b) phonon scattering rates.}
    \label{phT}
\end{figure}

For phonon relaxation, the coupling factor is computed as $G_{p p, i}=\frac{C_{p, i}}{\tau_{i}}$. The mode contribution specific heat $C_{p,i}$ can be evaluated by~\cite{shengbte}

\begin{equation}
C_{p,i}(T) = \frac{k_{\mathrm{B}}}{\Omega N} \sum_{i}\left(\frac{\hbar \omega}{k_{\mathrm{B}} T}\right)^{2} n_{0}\left(n_{0}+1\right),
\end{equation}
where $\omega$ is the phonon frequency, $\Omega$ is the BZ volume, $N$ is the number of sampling grid, and $n_0$ is the occupation number of phonons at equilibrium. The summation is over phonons in group $i$. In the above equation, temperature dependence is on the temperature of the current group $T = T_{p,i}$. The phonon relaxation rate $\tau_i^{-1}$ can be evaluated by the averaged three-phonon scattering rates in that subgroup at the RTA level since we have defined a lattice reservoir. We note that a prior work without lattice reservoir used empirical estimation here~\cite{MgB22020prl}. For one phonon mode $\lambda$, its scattering rate 
involving two other phonons ($\lambda',\lambda''$) and phonon absorption (+) and emission (-) processes can be computed by Fermi's golden rule:

\begin{equation}
\frac{1}{\tau_{\lambda}}=\frac{1}{N}\left(\sum_{\lambda^{\prime} \lambda^{\prime \prime}}^{(+)} \Gamma_{\lambda \lambda^{\prime} \lambda^{\prime \prime}}^{(+)}+\sum_{\lambda^{\prime} \lambda^{\prime \prime}}^{(-)} \frac{1}{2} \Gamma_{\lambda \lambda^{\prime} \lambda^{\prime \prime}}^{(-)}\right),
\end{equation}
\begin{equation}
\Gamma_{\lambda \lambda^{\prime} \lambda^{\prime \prime}}^{(+)}=\frac{\hbar \pi}{4} \frac{n_{\lambda^{\prime}}^{0}-n_{\lambda^{\prime \prime}}^{0}}{\omega_{\lambda} \omega_{\lambda^{\prime}} \omega_{\lambda^{\prime \prime}}}\left|V_{\lambda \lambda^{\prime} \lambda^{\prime \prime}}^{(+)}\right|^{2} \delta\left(\omega_{\lambda}+\omega_{\lambda^{\prime}}-\omega_{\lambda^{\prime \prime}}\right),
\end{equation}
\begin{equation}
\Gamma_{\lambda \lambda^{\prime} \lambda^{\prime \prime}}^{(-)}=\frac{\hbar \pi}{4} \frac{n_{\lambda^{\prime}}^{0}+n_{\lambda^{\prime \prime}}^{0}+1}{\omega_{\lambda} \omega_{\lambda^{\prime}} \omega_{\lambda^{\prime \prime}}}\left|V_{\lambda \lambda^{\prime} \lambda^{\prime \prime}}^{(-)}\right|^{2} \delta\left(\omega_{\lambda}-\omega_{\lambda^{\prime}}-\omega_{\lambda^{\prime \prime}}\right),
\end{equation}
where $V_{\lambda \lambda^{\prime} \lambda^{\prime \prime}}^{(\pm)}$ is the transition probability matrices elements and $\delta$ is the Dirac function. The temperature dependence of scattering rate $\tau_i^{-1}$ is then implicitly included in the occupation number of the interacting phonon modes $n_{\lambda^{\prime}}^{0},n_{\lambda^{\prime \prime}}^{0}$. This calculation then yields that $\tau_i^{-1}(T)$ should depend on the lattice bath temperature, or $T = T_{\rm lat}$. Based on these discussions, we can compute temperature-dependent phonon properties and find that $C_{p,i}$ has weak temperature dependence in the range of 300~K to 800~K (see Fig.~\ref{phT}(a)) and $\tau_i^{-1}$ is linearly related to lattice temperature (see Fig.~\ref{phT}(b)). In this study, we can neglect $C_{p,i}(T)$ and write the coupling factor as $G_{p p, i}(T)=\frac{C_{p, i}}{\tau_{i}(T_{\rm lat})}$, which is a function of lattice temperature.

\section{Finite-element modeling of heat transfer in an axisymmetric system}

In accordance with typical micro-Raman experiments, we create an axisymmetric system with sample suspended over a circular hole and two ends supported by Au/SiO$_2$ substrate. In such case, our extended MTM should be solved in 2D real space using finite-element method. We choose the mathematical module in a commercial solver COMSOL Multiphysics~\cite{comsol}. 

Since we divide the carriers into five representative subgroups, the system of partial differential equations describing the temperature field is of five dimensions with five independent variables $\mathbf{u}=[T_1,T_2,T_3,T_4,T_5]^{T}$ and the derivative operator $\nabla=[\frac{\partial}{\partial r},\frac{\partial}{\partial z}]$. In COMSOL, this is realized by the Coefficient Form of PDE solver. The reference temperature is set as 300~K and we have applied Dirichlet boundary conditions to this real space heat transfer problem. This includes prescribed temperatures of 300~K for all subgroups at the outer radial boundary, i.e., the system reaches thermal equilibrium at the boundary. In the case of a supporting substrate, we apply an outward flux term to acoustic phonon group.

\begin{figure}[h]
    \centering
    \includegraphics[width=6.5in]{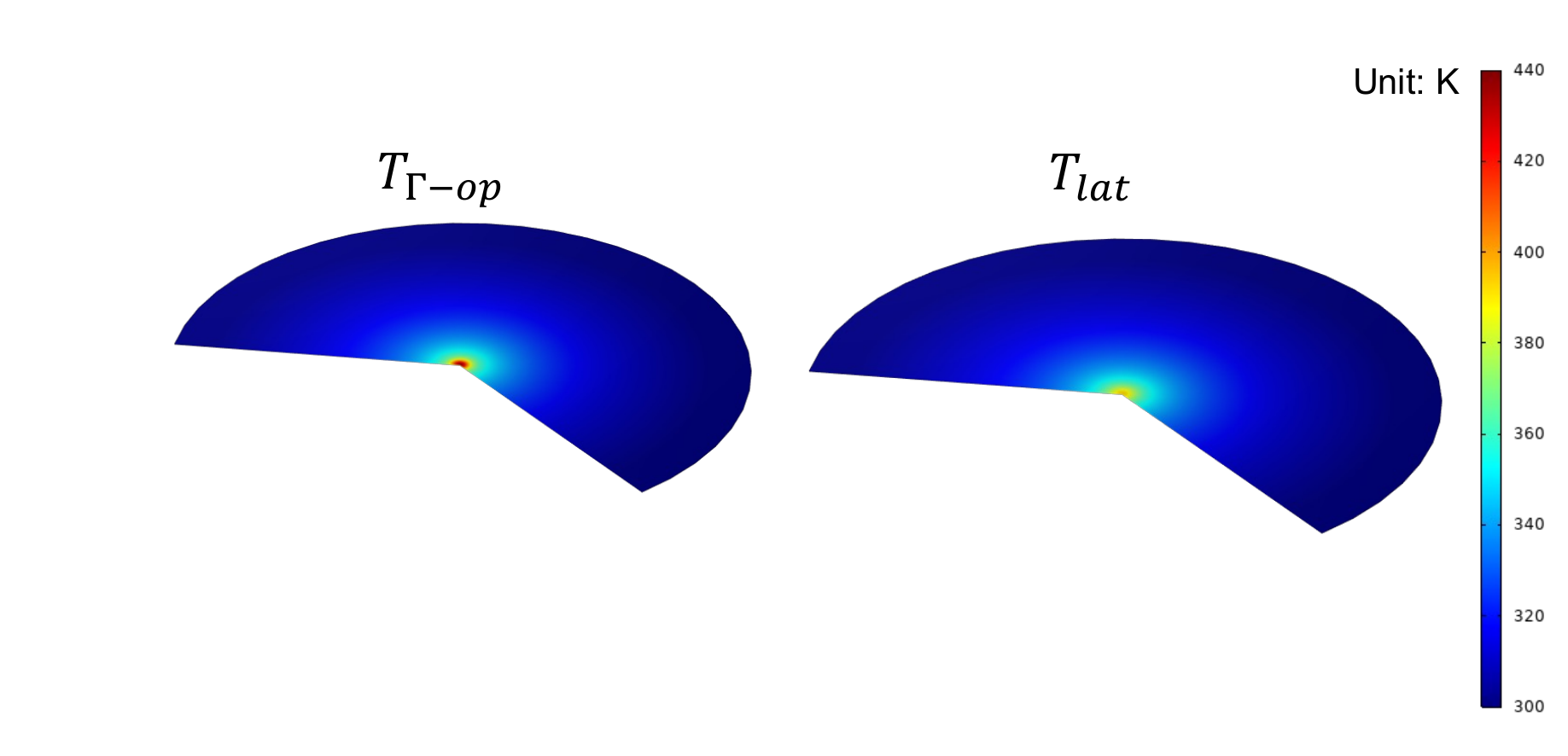}
    \caption{Simulated temperature profile $T(r,z)$ (unit: K) for zone-center optical phonons and lattice bath. Sample is suspended and has a thickness of 72~nm.}
    \label{comsolTrz}
\end{figure}

To compare with Raman experiment, one needs to average the temperature field to get experimentally probed Raman temperatures~\cite{laserprl2016}. For a temperature field $T(r,z)$, if we assume a Gaussian laser beam the weighted Gaussian average temperature $\langle T\rangle$ reads

\begin{equation}
\langle T\rangle=\frac{\displaystyle\int_{0}^{\infty} d z \int_{0}^{\infty} r d r \ T(r,z) Q(r, z)}{\displaystyle\int_{0}^{\infty} d z \int_{0}^{\infty} r d r\ Q(r, z)}
\end{equation}
where in the main text we define $Q(r,z)=\alpha I(r,z)$ with $I(r,z)$ being the laser intensity profile. The laser beam divergence relation $w(z)$ mentioned in the main text is calculated as $w(z) = w(0)\sqrt{1+(z/\beta)^2}$ with $w(0)$ being the laser spot size in radius and $\beta = \pi w(0)^2 n \lambda$, where $n$ is the indexes of refraction and $\lambda$ is the laser wavelength~\cite{peterExp}. In this work, we use the experimentally determined optical parameters to better reflect actual energy inflow to the system. In particular we have absorption coefficients $\alpha=4.35\times10^5$~cm$^{-1}$, reflectance $R=0.67$, refractive index $n=4.77$, laser wavelength $\lambda=532$~nm. The power level and laser spot size are reported in Ref.~\cite{peterExp}. The simulated temperature profile in such a radial system is shown in Fig.~\ref{comsolTrz} when laser power is $P=2.35$~mW and spot size is $r=\SI{0.71}{\micro\meter}$. The Gaussian averaged temperatures for zone-center phonons and the lattice in this case are 381~K and 392~K, respectively.

%\clearpage
\bibliography{Reference}% Produces the bibliography via BibTeX.